# Hindered surface diffusion of bonded molecular clusters mediated by surface defects

William S. Huxter,[1,*] Chandra Veer Singh,[1,2] and Jun Nogami[1,†]

[1] Department of Materials Science and Engineering, University of Toronto, 184 College Street, Toronto, Ontario M5S 3E4, Canada

[2] Department of Mechanical and Industrial Engineering, University of Toronto, 5 King's College Road, Toronto, Ontario M5S 3G8, Canada

[*] Present address: Department of Physics, ETH Zurich, Otto Stern Weg 1, 8093 Zurich, Switzerland.

[†] Corresponding author. Email address: jun.nogami@utoronto.ca

ABSTRACT

The design of low dimensional materials through surface assisted self-assembly requires a better understanding of the factors that limit and control surface diffusion. We reveal how substrate surface defects hinder the mobility of sub-monolayer organic adsorbates on a metal surface with the model CuPc/Cu(111) system. Post-deposition annealing bonds CuPc molecules into dendritelike clusters that are often mobile at room temperature. Surface defects on Cu(111) create energetic barriers that prevent CuPc cluster motion on the metal surface. This phenomenon was unveiled by the motion of small clusters that show rigid-body diffusion solely in the available space in between defects. When clusters are sufficiently surrounded by defects, they become completely pinned in place and become immobilized.

KEYWORDS: surface defects, on-surface synthesis, CuPc, Cu(111), scanning tunneling microscopy, density functional theory

I.   Introduction

The synthesis of low dimensional materials from molecular building blocks is a rich and promising area of molecular nanotechnology [1,2]. A common route for material synthesis involves facilitating covalent bonding between precursor molecules on an atomically flat substrate through an annealing treatment. The substrate, often a transition metal, assists in the reaction by limiting the adsorbed molecules to the two-dimensional (2D) surface and may act as a catalyst by providing adatoms for reaction intermediates [3-5]. One practical challenge with these coupled self-assemblies lies in determining the fundamental mechanisms and interactions that are important in the design of the formed polymer or oligomer. Discovering what controls these factors may lead to new physical and chemical insights in low dimensional systems. For instance, it has been demonstrated that polymer/oligomer morphology is greatly affected by changes in adsorbate surface mobility [6,7], temperature-dependent bonding mechanisms [8,9], and substrate temperature during precursor deposition [10].

Point defects on the substrate surface are often overlooked in this synthesis process. While experimental preparation techniques have advanced to generate nearly ideal substrates, surface defects are almost always unavoidable. These defects may physically and chemically interfere with precursors and modify reaction mechanisms. Thus, these defects merit a careful investigation. In this paper, we investigate the role of surface defects through the study of annealed sub-monolayer (ML) copper-phthalocyanine (CuPc) on Cu(111) with scanning tunneling microscopy (STM) and density functional theory (DFT) stimulations.

CuPc/Cu(111) is a model 2D π-conjugated molecule-metal system. Such systems form a variety of 2D structures [11,12] and have applications ranging from catalysis [13] to molecular spintronics [14,15] and organic electronic devices [16,17]. Specifically, the CuPc/Cu(111) system has been extensively characterized across many techniques from sub-ML to multi-layer coverage across a large temperature range [18-30]. Room temperature (RT) STM experiments show that CuPc is highly mobile on noble metal (111) surfaces at coverages under 1 ML [21,28,29,31]. The STM tip measures the time averaged motion of CuPc across the surface as a diffusive background feature, as well as interference patterns from CuPc scattering around surface defects and step edges. Similar metal-phthalocyanine (MePc)/metal systems [32-34] show evidence of C-C bond formation after annealing, however bond formation through annealing has yet to be investigated on CuPc/Cu(111).

We find that annealing CuPc on Cu(111) yields dendritelike clusters as a result of a dehydrogenation reaction that creates biphenyl links between molecules. CuPc appears to become immobilized after forming clusters, however, many smaller clusters diffuse and rotate on the Cu(111) terraces. Experimental observations and theoretical calculations demonstrate that clusters are immobilized by the energetic barriers created around surface defects that prevent CuPc diffusion. This reveals that the stochastic nature of surface defects can severely limit the cluster's surface mobility. Considering the significance of CuPc/Cu(111) as a prototypical system, we believe that these findings greatly add to the fundamental understanding of on-surface synthesis and may promote further studies of phthalocyanines, porphyrins, and other 2D π-conjugated molecules.

II. Experimental and Theoretical Details

Experiments were carried out in an RT ultra-high vacuum (UHV) STM chamber with a base pressure of ~$3.0 \times 10^{-10}$ Torr. Sample cleaning is described in Ref. [35]. CuPc was evaporated from a sublimated purified CuPc powder inside a direct-current heated quartz Knudsen cell. Current controlled deposition was monitored by an in-situ quartz crystal microbalance (QCM) and the deposition rate was measured to be ~0.15 ML/min. Post-deposition annealing occurred at ~573 K via a ceramic radiative heater placed on the back of the Cu(111) sample plate. This temperature was slightly higher than previously reported experiments on CuPc/Cu(111) that did not show bonding [22,23]. Postdeposition annealing times varied from 15 to 30 min. Images were collected with Pt-Ir tips in the constant current mode and post-processed with WSxM [36].

DFT simulations were performed using VASP [37] with projector augmented-wave potentials [38,39], generalized gradient approximation (GGA)-Perdew-Burke-Ernzerhof functionals [40], and DFT-D3 and Becke-Jonson damping for van der Waals corrections [41,42]. Simulation details included a kinetic energy cut-off of 500 eV and gamma point sampling (which is adequate due to the large super cells). Self-consistent calculations utilized a threshold of $10^{-4}$ eV/Å for force convergence and a threshold of $10^{-5}$ eV for total energy convergence. Four different Cu(111) slabs were created for simulations with different CuPc/Cu(111) and CuPc-CuPc/Cu(111) geometries and interaction schemes. Slab details are included in Ref. [35], including a discussion about the electronic structure of CuPc and comparison with other GGA simulations [43]. STM simulations followed the Tersoff-Hamann theory [44].

III. Results and Discussion

An example of covalently bonded CuPc clusters created at 0.25 ML is shown in Fig. 1(a). These dendritelike clusters are heavily branched, often extend across entire Cu(111) terraces, and are on the order of 50 CuPc molecules in size when not limited by step edges. Clusters are primarily formed out of two different intermolecular bonding orientations; a parallel orientation, shown in Fig. 1(b), where bonding lobes of the CuPc lie along parallel lines and an angular orientation, shown in Fig. 1(c), where bonding lobes lie 120° apart. DFT simulations of these bonding arrangements are shown in Fig. 1(d)-1(g) and they strongly support our observation of C-C bond formation. STM simulations and Cu-Cu distances from DFT calculations match the experimental observations well. Additionally, the reduced $C_{2v}$ symmetry of the adsorbed CuPc molecule is clearly visible. This symmetry is induced through interaction of the Cu(111) surface and can be observed as the bright and dark lobes seen on the CuPc molecules. $C_{2v}$ symmetries of MePc molecules are consistent with previous observations across (111) metal surfaces [19,30,45]. Prior theoretical calculations [46,47] have shown that brighter lobes align with close packed directions of the substrate. While the bonding lobes in Fig. 1(b) and the nonbonding lobes in Fig. 1(c) are slightly brighter, the opposite cases were also observed, albeit less frequently [35]. This suggests that CuPc orientation on the substrate has a negligible influence on the type of bond formed.

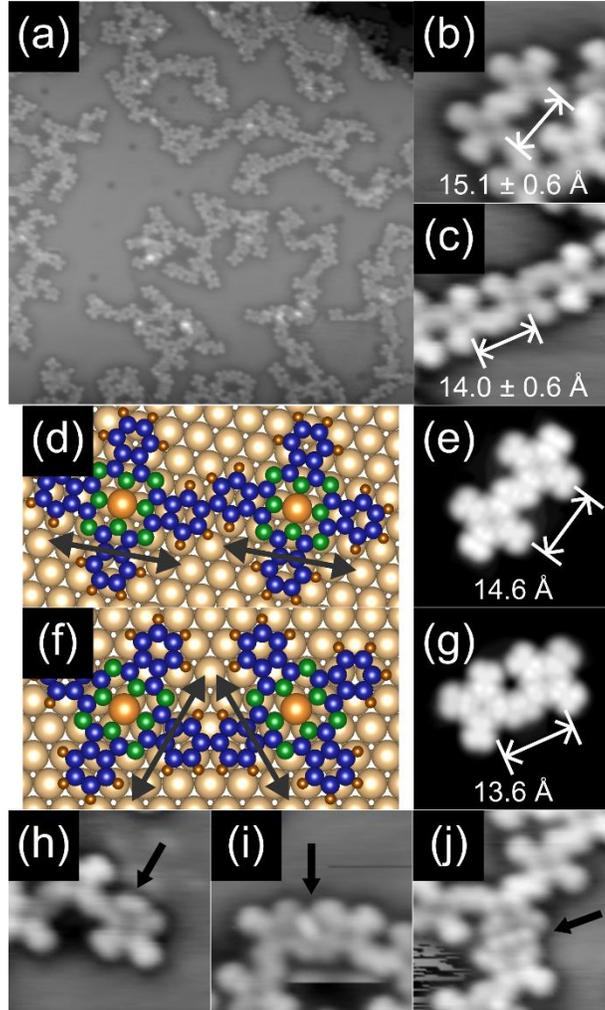

FIG. 1. (a) Annealed 0.25-ML CuPc on Cu(111) (500×500 Å$^2$). Specific parallel (b) and angular (c) bonding orientations (50×50 Å$^2$). Atomic model (d) and STM simulation (e) of parallel bonding. Atomic model (f) and STM simulation (g) of angular bonding. (h)-(j) Different CuPc-Cu adatom coordination structures (50×50 Å$^2$). STM simulations at +0.9 V. Imaged +0.9 V and 0.5 nA.

While each isoindole lobe of the CuPc molecule features two bonding sites on the peripheral carbon atoms, steric hindrance limits each lobe to one bond only. These bonding orientations match the structures observed previously on CuPc/Ag(111) [32], FePc/Cu(111) [33], and ZnPc/Cu(100) [34] (parallel bonding only) which suggests that the formation of the biphenyl link is minimally dependent on the choice of metal center and metal substrate. The biphenyl link bears a similarity to, but is different from, the bonding observed with annealed octaethyl-tetra-aza-porphyrin (OETAP) on Au(111) [10]. Post-deposition annealing transforms OETAP into phthalocyanine molecules which become bonded together through naphthalene links. These naphthalene links place the molecule centers closer together than biphenyl links and are inconsistent with our observations. Additionally, the benzene terminated lobes of CuPc are structurally and chemically different from the ethylene terminated lobes of OETAP which are required for creating the naphthalene links. Interestingly, the Au(111) surface reconstruction appears to spatially confine the annealed OETAP clusters which preferentially grow in the FCC regions of the herringbone structure [10]. On Cu(111) the CuPc clusters show no preferred morphology and tend to grow in all directions.

In rare instances, CuPc-Cu adatom coordination was observed. Several seemingly hierarchical structures were formed as depicted in Fig. 1(h)-1(j). The lobes near Cu adatoms are adsorbed closer to the surface as they appear darker in STM images. It is likely that CuPc are stabilized by this interaction, similar to how single Ag adatoms have been shown to stabilize CuPc on Ag(100) [48] and to how other adsorbed molecules interact with Cu adatoms on Cu(111) and Cu(100) [49-56] . We do not believe this coordination is caused by the hydrogen lost during the dehydrogenation reaction since both atomic and molecular hydrogen would not stick to or interact with the Cu(111) surface within the relevant temperature range [57,58].

Different CuPc coverages were studied to measure changes in cluster size upon annealing. At 0.15 ML, many smaller clusters were observed (compared to 0.25 ML) that were spread evenly over the surface. The smallest resolvable cluster contained three CuPc molecules. Surprisingly, repeated STM scans (~10 min per image) revealed rigid-body motion by a fraction of the clusters, including diffusion and rotations, as shown in Fig. 2. Cluster diffusion was observed to be as large as a several nanometers in between scans. Rotations, which were rarely observed, featured stable positions on the Cu surface that were roughly 60° apart [see Fig. 2(a) to 2(e)]. This suggests that rotating clusters were aligning with close packed directions. When mobile clusters interacted with immobile clusters, as in Fig. 2(f)-2(h), the CuPc-CuPc distances from the nearest interacting CuPc molecules were sufficiently large to conclude that they were not bonding to each other. Clusters bonded to CuPc at step edge as well as clusters with CuPc-Cu adatom coordination were observed to be immobile.

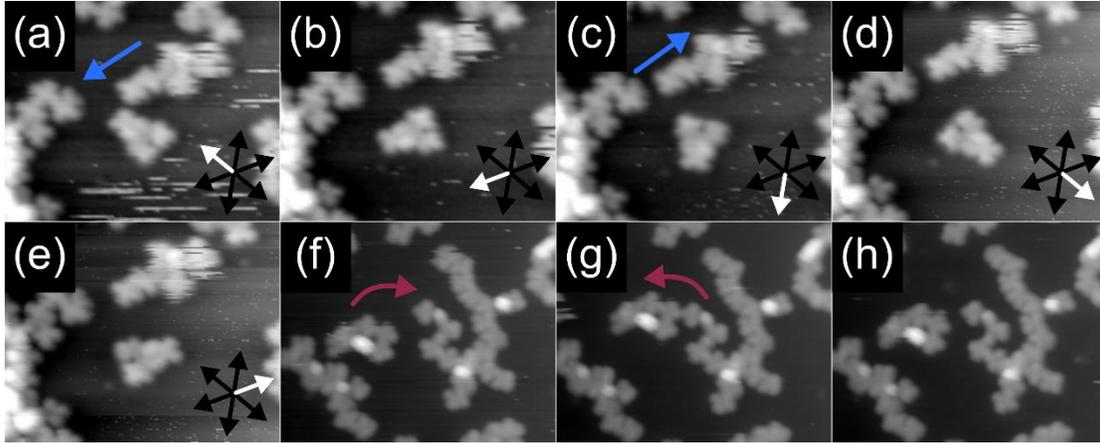

FIG. 2. 0.15-ML CuPc on Cu(111) with post-deposition annealing. (a)-(e) Rotational configurations of a CuPc cluster (-1.0 V, 0.5 nA, 200×160 Å$^2$). The stable positions are indicated by white and black arrows. Diffusion back and forth between different orientations was observed across 17 sequential images recorded over a 3.5-h period. Translations by another cluster are also indicated by blue arrows. (f)-(h) Interaction between different CuPc clusters across three sequential scans highlighted with red arrows (-1.1 V, 0.5 nA, 250×200 Å$^2$).

The observed cluster motion was varied and complex. Mobile clusters were sometimes well-resolved and other times noisy from scan to scan. Evidently, mobile clusters could be motionless or in motion while the STM tip was scanning directly above the cluster. In general, smaller clusters were more likely to be mobile. One possible explanation is that larger clusters diffuse at slower rates due to their size and eventually, at a threshold size, become too large to move. However, this does not explain why some very small clusters (< 5 CuPc) were immobile yet many larger clusters (up to ~20 CuPc) were mobile. Possibilities of tip induced effects, such as a chemisorption mechanism [26,59-61] or local electric fields [62] are not likely as motion was observed at both scanning biases and no voltage pulses were applied from the scanning tip.

It is already understood that CuPc molecules on clean Cu(111) are highly mobile at RT at sub-ML coverages, and do not appear in STM images at room temperature. At coverages approaching 1 ML, the presence of mobile CuPc becomes apparent as a smooth background that is higher than the bare substrate [31,63]. An ordered 2D CuPc phase only becomes apparent to STM as the coverage reaches 1 ML and the second layer of CuPc only begins to form once the first ML is completed [27]. Here, we additionally report that bonded CuPc molecules also remain mobile on the surface and that their diffusive behavior deviates from that of their nonbonded counterparts.

To further understand cluster mobility, additional CuPc can be deposited on a surface with CuPc clusters without subsequent annealing. Thus, differences in mobility between clusters and single molecules is easily distinguishable in the same STM scan. Fig. 3 displays the annealed 0.15-ML CuPc surface before and after adding an additional 0.15=ML CuPc without annealing. In Fig. 3(a) noisy streaks, which are topographically the same height as the immobile clusters, are indicated with white arrows. The diffuse background created by highly mobile CuPc in Fig. 3(b) sits at a lower topographic height and is considerably different than the noisy streaks in both STM scans (see Refs. [31,63] for more information about the diffusive background). We attribute the noisy streaks to CuPc clusters – possibly bonded CuPc pairs – that were moving too quickly to be fully resolved by the tip. These streaks appear brighter than the diffusive background because they are diffusing more slowly than individual CuPc molecules and therefore spend a larger amount of time under the STM tip.

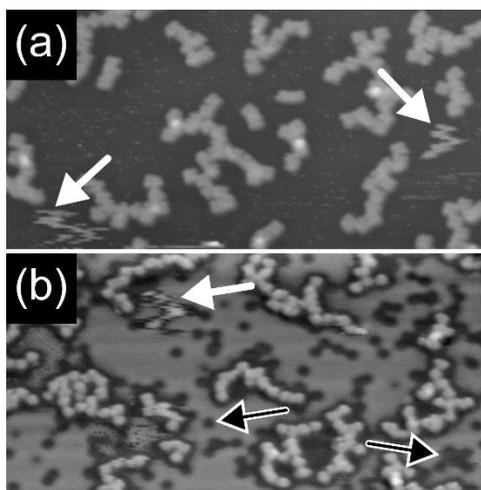

FIG. 3. Addition of 0.15-ML CuPc without annealing. (a) STM scan of annealed CuPc before additional deposition (-2.0 V, 0.1 nA, 600×350 Å$^2$). (b) STM scan after additional deposition of CuPc (-1.2 V, 0.5 nA, 600×350 Å$^2$). White arrows indicate mobile clusters and black arrows indicate substrate surface defects.

Single CuPc molecules also scatter off CuPc clusters. Regions devoid of the diffusive background around surface defects are also found around clusters. A few surface defects are indicated in Fig. 3(b) with black arrows. Regions between defects that CuPc molecules are physically incapable of occupying show no diffusive background, giving the appearance of extended defect structures (more details in Ref. [35]).

In Figs. 2 and 3 we have observed two types of motion: (1) highly mobile single CuPc molecules that appear as a diffusive background, and )2) bonded CuPc clusters that can either be unresolvable in a single image (streaky) or fully resolved from image to image. While the inability to change the operating temperature of the STM limits our ability to measure diffusion parameters (diffusion prefactor $D_0$ and activation energy $E_A$) we can estimate and compare the diffusion speeds of CuPc molecules and CuPc clusters (calculation details in Ref. [35]). Additionally, measured displacements can lead to estimates for the diffusion coefficient D [$D = D_0 \exp(-E_A/k_BT)$] at RT. Single CuPc molecules must be moving at speeds at least on the order of $1.5 \times 10^3$ nm/s with D ~ $10^{-11}$ cm$^2$/s, given our tip dwell time and the inability to resolve any molecular features. The clusters, however, must be moving significantly slower. Clusters that produce noisy streaks (such as the streaks denoted by white arrows in Fig. 3) can be moving as slowly as $1.2 \times 10^0$ nm/s and clusters that only appear to be moving in between scans (such as the clusters in Fig. 2) can be moving as slow as $10^{-2}$ to $10^{-3}$ nm/s. These values correspond to D ~ $10^{-14}$ cm$^2$/s to $10^{-17}$ cm$^2$/s, which is three to six orders of magnitude smaller. It should be noted that these estimates are only lower bounds and that it is possible that some clusters are moving faster. It is difficult to observe motion on the order $10^1$-$10^2$ nm/s as that is comparable to our tip scanning speed. We also find that there is a weak correlation between the size of the cluster and the motion; larger clusters appear to move more slowly than smaller clusters.

A series of DFT simulations provides insights into the behavior of sub-ML CuPc on Cu(111) by measuring the change in CuPc adsorption energy across a variety of CuPc/substrate interactions. Figure 4 displays the clear trend that CuPc is further stabilized by increasing the interaction with substrate atoms and is less stable upon increasing CuPc-CuPc interaction. A full description of the simulation details are found in Ref. [35]. When CuPc interacts at or near a step edge the adsorption energy is ~1 eV more stable. Isolated CuPc and monolayer CuPc occupy a range of roughly 0.48 eV due to the different possible positions of the molecules on the Cu(111) surface. This value represents an upper bound for the diffusion barrier of CuPc on Cu(111) [35]. The adsorption energy begins to quickly increase as CuPc lobes are removed from the Cu(111) surface and forced to interact with other CuPc. CuPc in the second ML are ~2.5 eV less stable than in the first ML.

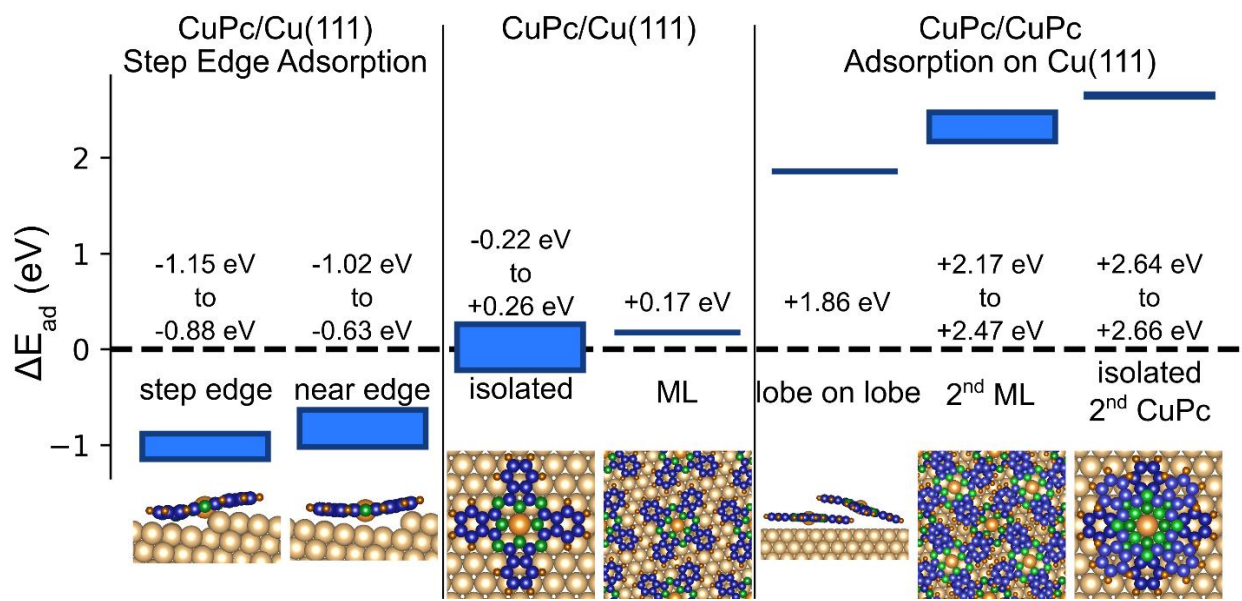

FIG. 4. Relative stability of CuPc across different interaction schemes. CuPc adsorption on Cu(111) is used as the reference energy. The energy ranges come from taking the range of adsorption energies from several different geometries (details in Ref. [35]). Negative energies are more stable.

These energies provide results consistent with the experimental observations. CuPc deposited onto Cu(111) at RT can further increase its stability by adsorbing at step edges [which has been previously observed on CuPc/Cu(111) [26,27], CuPc/Au(111) [64], and FePc/Cu(111) [65]]. The ~1-eV increase in stability is larger than the thermal energy of CuPc as the molecules do not desorb from the step edges. When CuPc is deposited onto the post-annealed surface (with bonded CuPc clusters), no single CuPc molecules remain on top of clusters. If CuPc were to land on top of a cluster, the CuPc would increase its energetic stability by transitioning off the cluster and onto the available Cu(111) surface.

Repeated STM scans of the cluster plus single molecule surface, shown in Fig. 5, clearly show that the highlighted surface defects hinder cluster mobility The clusters surrounded by defects outlined in white and red move within their defect free region from scan to scan and never move over any defects. The cluster surrounded by defects outlined in blue moves very quickly inside its defect free region, then appears to escape that space in Fig. 5(b) as the noisy streaks indicative of the cluster disappears. It is likely that this cluster moved through the two leftmost highlighted defects as a characteristic noisy streak is observed to the left of those defects in Fig. 5(c). The same type of behavior is also shown with the mobile cluster indicated by the white arrow in Fig. 3(b).

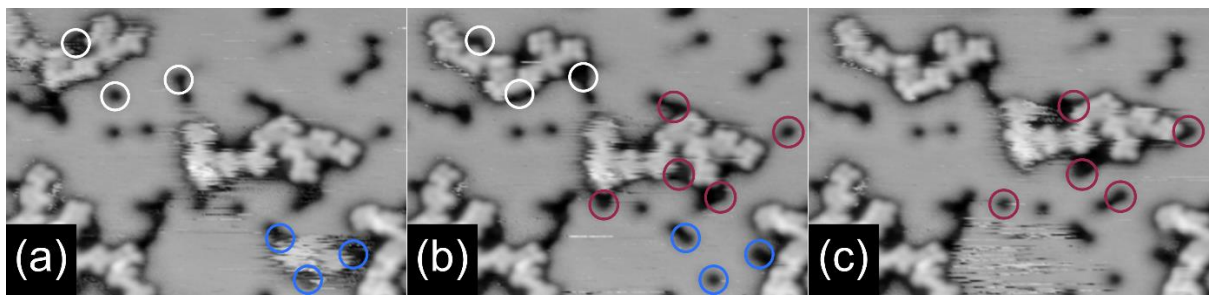

FIG. 5. (a)-(c) Sequential scans that track mobile cluster movement. From panel (a) to (b) the topmost cluster moves until it is pinned by surface defects outlined in white. Another cluster, enclosed by defects outlined with blue, disappears in panel (b). From panel (b) to (c) motion of another cluster is hindered by defects outlined in red. All imaged at -1.0 V, 0.4 nA, 300×180 Å$^2$.

Surface defects acting as pinning sites for CuPc cluster motion provide an adequate explanation for the observed cluster motion. Very small clusters should almost always be mobile as they would be able to move between most surface defects. If a cluster is surrounded by a few defects, such as those surrounded by highlighted defects in Fig. 5, it is possible that motion is restricted to a small area. If surrounded by enough surface defects, the entire cluster becomes immobilized. As the cluster size increases, such as when annealing at higher CuPc coverages, it is more likely that clusters are surrounded by enough surface defects to become immobilized. Due to the stochastic nature of surface defects, it is possible that some larger clusters remain mobile while smaller clusters are immobile. It is also likely that larger clusters naturally diffuse more slowly on the surface. However, surface defects, step edges, and Cu adatom coordination appear to be the most important factors responsible for cluster immobilization. The importance of surface defects revealed here may have significant consequences across a large variety of molecule/metal systems. This mechanism may explain the mobility (and lack of mobility) in other cases of rigid-body diffusion of bonded molecules [10,66], especially those where the mobility of the molecular cluster largely deviates from that of individual molecules.

Auger spectroscopy on Cu(111) single crystals prepared in a similar manner to our samples has revealed that sulfur is likely the primary surface contaminant [67]. Annealing cycles used in sample preparation enables sulfur to diffuse from the bulk to the surface. This specific phenomenon has been used previously to study S adatoms on Cu [68,69]. Figures 6(a) and 6(b) display typical STM images of surface defects on Cu(111) before and after a tip change. The observed morphology is characteristic of STM images of single sulfur atoms on noble metal surfaces. Specifically, the protrusions imaged in Fig. 6(a) have a height of ~0.016 nm and a full width at half max of ~0.36 nm, similar to the values of isolated S on Cu(100) [70] and S on Au(111) [71]. The imaged defects do not appear like the $Cu_2S_3$ complexes or sulfur induced reconstructed steps observed on Cu(111) [72,73]. Given the appearance of the defects in our data, we assume that isolated sulfur atoms are the primary surface defect responsible for hindering cluster motion.

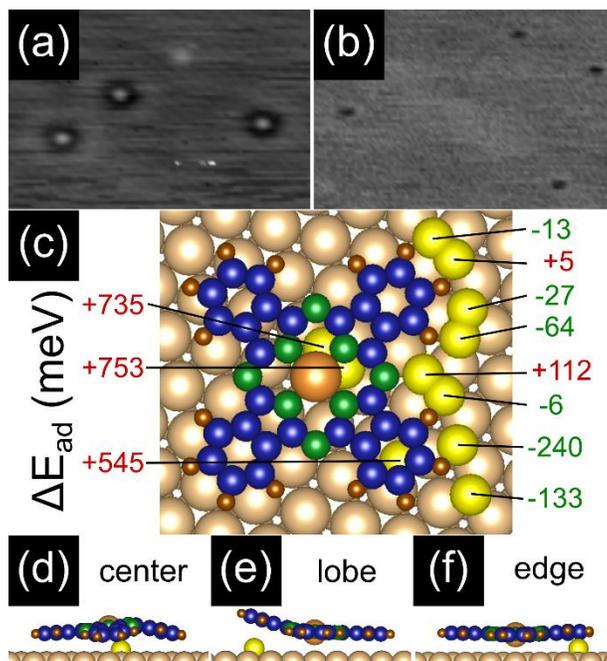

FIG. 6. Surface defects on Cu(111) before (a) and after (b) a tip change (-1,2 V, 0.2 nA, 120×80 Å$^2$) (c) Relative adsorption energies of CuPc near and on top of S defects. (d)-(f) Three relaxed geometries of CuPc by S on Cu(111).

DFT simulations of this type of defect reveal how cluster motion is hindered. Eleven different FCC and HCP sites were populated with a sulfur atom either underneath or beside an isolated CuPc molecule. The locations of these sites relative to the adsorbed CuPc and the changes in adsorption energy, $\Delta E_{ad}$, are shown in Fig. 6(c). Positive values indicate a less favorable adsorption configuration. Figures 6(d)-6(f) display the relaxed geometry of a few cases. Sulfur adatoms placed near the edge of the lobes show a slight decrease in adsorption energy. However, these small changes are mostly comparable to changes in energy as CuPc moves across the Cu(111) surface and activation barriers for MePc diffusion on Ag(100) [74,75] and on Au(111) [76]. Thus, sulfur near the edges of CuPc minimally affects the molecule. Sulfur underneath CuPc however, is not favorable as it interferes with the adsorbate-substrate interaction. The unfavorable adsorption of CuPc on top of sulfur demonstrates that CuPc molecules (and clusters) experience a significant energetic barrier as they try to move over a surface defect. In the case of single molecules, this produces scattering patterns [26,27]. In the case of bonded CuPc clusters, this hinders motion in the xy-plane and can effectively trap clusters.

IV. Conclusion

A mechanism is presented whereby molecular clusters are immobilized by substrate surface defects. Dendritelike CuPc clusters reveal complex rigid-body mobility (and immobility) on the Cu(111) surface that is, to a first-order approximation, cluster size dependent but also subject to stochastic randomness from the distribution of substrate surface defects. Surface defect pinning, in combination with substrate step edges, native adatoms coordination, and size dependent diffusion constants provide a complete picture of how and why mobile precursor molecules may combine into immobile polymers and oligomers.

V. Acknowledgments

This research was possible through the support provided by Natural Sciences and Engineering Research Council of Canada (NSERC), Grants No. RGPIN-2017-06069 and No. RGPIN-2018-04642. Hart Professorship, and the University of Toronto. Computations were conducted through the Compute Canada facilities, namely the Niagara cluster.

# Supplemental Material

# Hindered surface diffusion of bonded molecular clusters mediated by surface defects

William S. Huxter[1,*], Chandra Veer Singh[1,2], Jun Nogami[1,†]

[1] Department of Materials Science and Engineering, University of Toronto, 184 College St., Toronto, ON M5S 3E4, Canada

[2] Department of Mechanical and Industrial Engineering, University of Toronto, 5 King's College Road, Toronto, Ontario, M5S 3G8, Canada

[*] Present address: Department of Physics, ETH Zurich, Otto Stern Weg 1, 8093 Zurich, Switzerland

[†] Corresponding Author: jun.nogami@utoronto.ca

## S1. Additional experimental details

The Cu(111) substrate was cleaned by alternating cycles of $Ar^+$ sputtering at 0.85 KeV and annealing at ~753 K. Sputtering cycles lasted 20 minutes and annealing cycles varied from 10 to 50 minutes. Cleaning cycles were repeated until STM images showed relatively large and clean terraces. We found that annealing our sample at ~753 K gave a surface defect concentration of roughly 0.0023 ML and annealing at ~ 853 K gave a defect concentration of roughly 0.0055 ML. We also found that the post-deposition annealing treatment of CuPc at ~573 K did not change the surface defect concentration. Surface defect concentrations were estimated by counting visible defects in STM scans and normalizing by the scan area. Measurements of the CuPc-CuPc distances shown in Fig. 1 were calibrated using the well-studied Cu(111)-$C_{60}$-p(4×4) structure to minimize measurement errors.

## S2. Additional theoretical details

In all DFT simulation supercells the bottom layer of Cu atoms where fixed in position and all others were free to relax. Additionally, a vacuum spacing of at least ~ 15 Å was used to eliminate effects of periodicity in the z direction. A 4-layer Cu(111) slab (10×10×4 atoms in a 25.6×22.1×25 $Å^3$ supercell) was used to model single CuPc adsorption and CuPc adsorption near a sulfur impurity. A 3-layer Cu(111) slab (15×10×3 atoms in a 38.3×22.1×25 Å$^3$ supercell) was used to model CuPc-CuPc bonding and non-bonding interactions. This slab was a larger version of the previous slab to accommodate a second adsorbed CuPc. A 4-layer Cu(111) slab based on the 1 ML CuPc/Cu(111) structure described by the superstructure matrix (6, 2 | 1 6) (experimentally determined by refs. [1] and [2]) was created to model the adsorption geometry of 1 ML of CuPc on Cu(111). This slab featured 34×4 atoms in a 13.9×13.2×28.4 Å$^3$ supercell. A 370-atom Cu(111) step edge slab (in a 25.6×20.7×25 Å$^3$ supercell) was created to model CuPc adsorption near step edges of the Cu(111) surface. This slab was a modified version of the 10×10×4 Cu atom slab. A tilt of ~5.77 degrees was applied to the initially flat Cu(111) surface to align the step edge with the periodicity of the slab – effectively creating a periodic step edge along a close packed direction in one axis. The applied tilt resulted in three of the four Cu(111) layers losing one row of atoms and thus the original 400-atom slab was turned into a 370-atom slab.

The following details the simulations displayed in Fig. 4 of the main text. Isolated CuPc adsorption was tested across 49 different positions inside of a 1×1 Cu(111) unit cell within the 10×10×4 Cu atom Cu(111) slab and additional position with different rotations to sample the energy changes a CuPc might experience as it diffuses around the surface. The 49 different positions (without rotations) featured electronic relaxation and no structural relaxation (done to save computation time and this is justifiable since the differences in energy were quite small). All other simulations included both electronic and structural relaxation. All the energies calculated are included in the energy range in Fig.4 for isolated absorption (-0.22 to 0.26 eV). The adsorption energy of CuPc in the 1ML structure also falls in this range. This energy range (~0.48 eV) provides an upper bound for the diffusion energy of the CuPc molecule on Cu(111). This claim is reason given the values for diffusion of CoPc on Ag(100) (0.14 to 0.19 eV barrier determined experimentally and 0.08 to 0.31 eV determined with simulations [3]) and CuPc on Ag(100) (~0.081 eV determined experimentally [4]).

Two different CuPc adsorption geometries were used to test adsorption on the step edge case, a single lobe extending across the step edge and two lobes extending across the step edge. Those position were chosen since most CuPc adsorbed on Cu(111) step edges feature a single lobe on the upper terrace, however adsorption with two lobes have also been observed [1,5]. During energetic relaxation the Cu center of the CuPc naturally relaxed to sit close to the Cu atoms at the edge of the upper terrace, similar to previous simulation studies of CuPc on Au(111) step edges [6] and FePc on Au(111) step edges [7]. Adsorption of CuPc near the step edge was performed in a similar manner, except that the CuPc was place further away from the step edge so that only a fraction of the lobe(s) extended to the upper terrace.

Simulations that remove CuPc-Cu(111) interaction in favor of CuPc-CuPc interaction were performed to give a rough idea of what the change in adsorption energy might be. Placing one lobe of a CuPc on top of another adsorbed CuPc already significantly increase the adsorption energy (Fig. 4 of the main text). Four different geometries were modelled to test different 2$^{nd}$ ML CuPc structures assuming that the second layer forms the exact same superstructure as the 1$^{st}$ ML. This assumption is incorrect as it has been reported that multilayer CuPc/Cu(111) exhibits a lattice contraction [1], however, the second layer only slightly deviated from the first. Two different CuPc geometries were used to simulate stacking of isolated CuPc. One placed all the of lobes on top of each other and the other was a 45-degree rotated version. There is no experimental evidence to indicate that this structure is stable, and this is reflected by the higher adsorption energies of these simulations.

In Fig. 4 of the main text, $\Delta E_{ad}$ was calculated as;

$$\Delta E_{ad} = \left(E_{CuPc/Substrate} - E_{free\ CuPc} - E_{Substrate}\right) - E_{Ad} \qquad (1)$$

where $E_{CuPc/Substrate}$ is the energy of the total system, $E_{free\ CuPc}$ is the energy of the free CuPc molecule, $E_{Substrate}$ is the energy of the substrate, and $E_{Ad}$ is the adsorption energy of CuPc on Cu(111). For each of the four different Cu(111) slabs a different $E_{Substrate}$ was used.

A similar calculation scheme was used for calculating the change in adsorption energy for the simulations carried out with sulfur substrate defects (in Fig. 6 of the main text). The change in adsorption energy was calculated as;

$$\Delta E_{ad} = \left(E_{CuPc/Cu(111)+S} - E_{free\ CuPc} - E_{Cu(111)+S}\right) - E_{Ad} \qquad (2)$$

where $E_{CuPc/Cu(111)+S}$ is the energy of the total system, $E_{free\ CuPc}$ is the energy of the free CuPc molecule, $E_{Cu(111)+S}$ is the energy of the substrate with the sulfur defect.

## S3. Bonding alignment on Cu(111), CuPc-Cu coordination, and cluster statistics

In the majority of parallel bonded CuPc the bonded lobe axis of the molecules aligned with a closed packed direction of the Cu(111) substrate. However, this type of alignment was not the only type of adsorbate-substrate alignment observed. Fig. S1(a) and S1(b) displays chains of CuPc bonded in the parallel orientation with differing bright lobes (indicated with white arrows) with respect to the bonding lobe. The difference in adsorbate-substrate alignment is clearly shown by the difference in lobe brightness. The majority of CuPc bonded in the angular orientation aligned the non-bonded lobe axis with a closed packed direction. However, angular bonded CuPc with the opposite adsorbate-substrate alignment was also observed. Both cases are shown in Fig. S1(c) and S1(d). STM measurements of Cu-Cu distances between pairs of bonded CuPc in the same orientation appeared independent of substrate alignments (no statistically significant difference). This is consistent with the notion that the adsorbate-substrate alignment, which arises due to van der Waals interaction, is a non-dominant interaction force that does not significantly affect the formation of the CuPc clusters. This is also consistent with the fact that CuPc form many bonds with varying orientations and adsorbate-substrate alignments. For example, the CuPc centered in Fig. S1(d) forms three CuPc-CuPc bonds – two in the angular orientation and one in the parallel orientation.

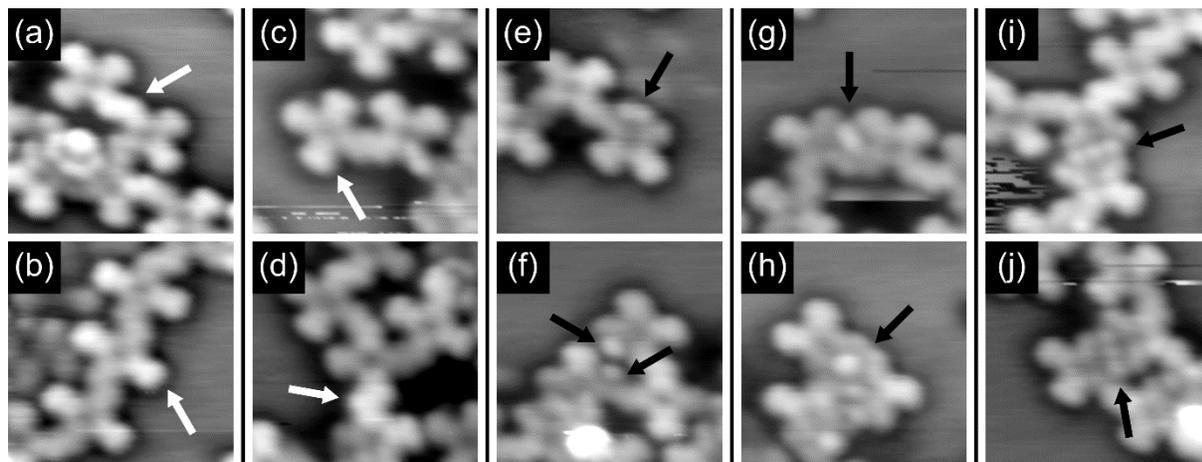

FIG. S1. (a-d) Images of CuPc clusters (at 0.25 ML) showing different types of alignment with respect to the Cu(111) substrate. (e-f) Isolated Cu coordination at the corner site of CuPc. (g-h) Cu coordination connecting two CuPc together. (i-j) Line of three Cu connecting two CuPc together. (a) and (c-j) imaged at +0.9 V and 0.5 nA. (b) imaged at +1.6 V and 0.5 nA. All panels are sized 50×50 Å$^2$.

Fig. S1(e) to S1(j) display rare configurations of CuPc in clusters. These CuPc form a variety of structures with what can be interpreted as Cu adatoms (indicted by black arrows). These Cu adatoms may have originally been found on the Cu(111) terraces or may have originated from step edges that were free from CuPc while the substrate was held at elevated temperatures. Several molecule-Cu adatom structures have been observed after post-deposition annealing on Cu(111) substrates [8-12] and other molecule-Cu adatom structures have been observed on Cu(100) without annealing [13-15]. While the Cu adatom features in Fig. S1 are clearly distinguishable and separate from the CuPc, it is not clear if each STM maxima is created by a single Cu adatom or several Cu adatoms.

The coordination between CuPc and the Cu adatoms appears to freeze the molecules in place. The lobes closest to the Cu adatom appear darker than the other lobes either due to the lobes being brought closer to the Cu(111) surface through bond formation or via a charge transfer mechanism. Additionally, the Cu adatom coordination with CuPc observed here is unique due to the variety (or hierarchy) amongst the coordinated structures. Cu adatoms have attached to the corner in between two lobes (Fig. S1(e) and S1(f)) where they most likely coordinate with the lone pair electrons on the nitrogen atom located there. The Cu adatoms have also coordinated in between two CuPc molecules through either a single Cu adatom site (Fig. S1(g) and S1(h)) or by forming a linear chain of three Cu adatom sites (Fig. S1(i) and S1(j)).

Clusters can be statistically described through by the number of bonds and the types of bonds for each CuPc. This is shown in Fig. S2 for measurements made across 460 molecules on the 0.25 ML covered surface. The majority of CuPc formed two bonds and almost never formed more than three bonds. Steric hindrance provides an adequate explanation. While CuPc can ideally form 16 different bonding configurations (4 lobes with 2 bonding sites and 2 bonding orientation) it was observed that only one bond forms per lobe. Due to the shape of the molecule an angular or parallel bond on one side of the lobe makes it impossible for another CuPc to bond on the other lobe. Additionally, since many small clusters form before larger clusters form (as seen on the 0.15ML CuPc covered surface) it is less likely that a doubly-bonded or a triply-bonded CuPc will be able to find another CuPc (or another CuPc cluster) that could find a way to fit and bond onto the available lobes.

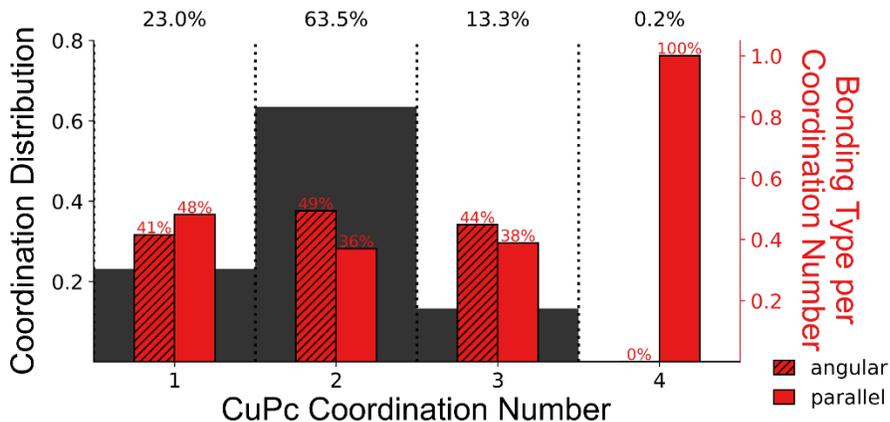

FIG . S2. CuPc cluster statistics of the 0.25 ML covered surface. Coordination number for CuPc in clusters and the distribution of their bonding type.

The ratio of angular bonding to parallel bonding appears to be very close to one-to-one. In the collected data 55% of identifiable CuPc-bonds were angular and 45% were parallel. There was a slight preference for parallel bonding on singly-bonded CuPc and a slight preference for angular bonding on doubly- and triply-bonded CuPc. Note that the percentages shown in Fig. S2 do not add up to 100% as some bonds were not clearly distinguishable as angular or parallel in nature. This was either due to imaging conditions (tip instability), CuPc bonding to other CuPc outside of the imaging window, or Cu adatom coordination.

## S4. Additional DFT simulations of cluster formation

In this section the formation of the C-C bond is investigated and characterized with DFT. This characterization includes STM simulations and electron localization function (ELF) plots across eight different CuPc-CuPc geometries (bonding and non-bonding schemes with two different molecular orientations and two different adsorbate-substrate alignments). Spin-polarized Projected Density of States (PDOS) simulations were also performed across several structures.

Fig. S3 displays the DFT relaxed atomic models with empty and filled state STM simulations. The computed Cu-Cu distances are also included on the far right. In all cases without dehydrogenation reactions the CuPc pairs are significantly further away from each other (beyond the experimentally measured range) and STM simulations show a plane between the molecules that is not experimentally observed. In the dehydrogenated simulations, the area directly above the newly formed C-C bond is not a minimum which matches experimental observations.

| Atomic Model | +0.9 V Empty State | -2 V Filled State | $d_{Cu-Cu}$ (Å) |
|---|---|---|---|
| angular | | | 13.6 |
| angular | | | 15.6 |
| alt. angular | | | 13.6 |
| alt. angular | | | 15.6 |
| parallel | | | 14.6 |
| parallel | | | 16.8 |
| alt. parallel | | | 14.6 |
| alt. parallel | | | 16.0 |

FIG. S3. CuPc-CuPc interaction with and without dehydrogenation reaction. Each orientation (angular and parallel) with its typical and alternative adsorbate-substrate alignment has relaxed with and with the removal of two hydrogens. Arrows on the atomic models show closed pack direction of Cu(111).

The STM simulations also show a 2-fold symmetry around each CuPc as a result of the adsorbate-substrate alignment, however the difference between bright and dark lobes are only faintly visible. In some cases only the bonded lobe is brighter and while most simulations slightly brighten the lobes aligned with the closed packed direction of the Cu(111) surface there is one exception – the alternative parallel bonded scheme. Those STM simulations show brighter lobes along the bonded axis instead of the Cu(111) close packed direction despite the fact that the non-bonding scheme follows the predicted behavior. This suggests that the influence of the bonded CuPc in the parallel orientation affects the molecule roughly on the same order as the van der Waals interaction.

ELF plots, shown in Fig. S4, provide a more direct method to look at the 2-fold symmetry of bonded and non-bonded CuPc. In Fig. S4 the central column displays a 2D ELF slice in the xy-plane across the average z-height of the molecule and the columns to the left and right shows slices below and above the average z-height. As expected, the center slice shows all bonded orbitals. The slices below and above the average z-height show slightly brighter (more localized) orbitals across opposite lobes that match the STM simulations. These plots also provide clearer 2-fold symmetries in comparison to the STM simulations.

Fig. S5 demonstrates that the bonding of the CuPc has little to no effect on the electronic properties of CuPc compared to non-bonding CuPc. The first PDOS plot (free CuPc) clearly displays the molecular orbitals. Orbitals near the Fermi level ($a_{1u}$, $b_{1g\uparrow}$, $b_{1g\downarrow}$, and $e_g$) are labeled for clarity and can be tracked throughout the other PDOS plots. Simulations of the isolate molecule and the 1 ML structure agree with previous PBE-GGA simulations [16-18]. While it is well-established that hybrid density functionals (such as HSE) are needed to correctly place the $b_{1g\uparrow}$ orbital lower in energy than the $a_{1u}$ orbital and place the $b_{1g\downarrow}$ orbital higher in energy than the $e_g$ orbital this was not practically possible due to simulation size and computational resources. Simulations with bonded CuPc appear very similar to each other with no significant changes. The biphenyl bond creates no apparent effect on electronic (and spin) properties. Both Cu atoms retain their spin ½ state. Given that CuPc clusters are rigid, highly disordered, and possess spin ½ states at their center, they present themselves as an interesting system that warrants further investigation.

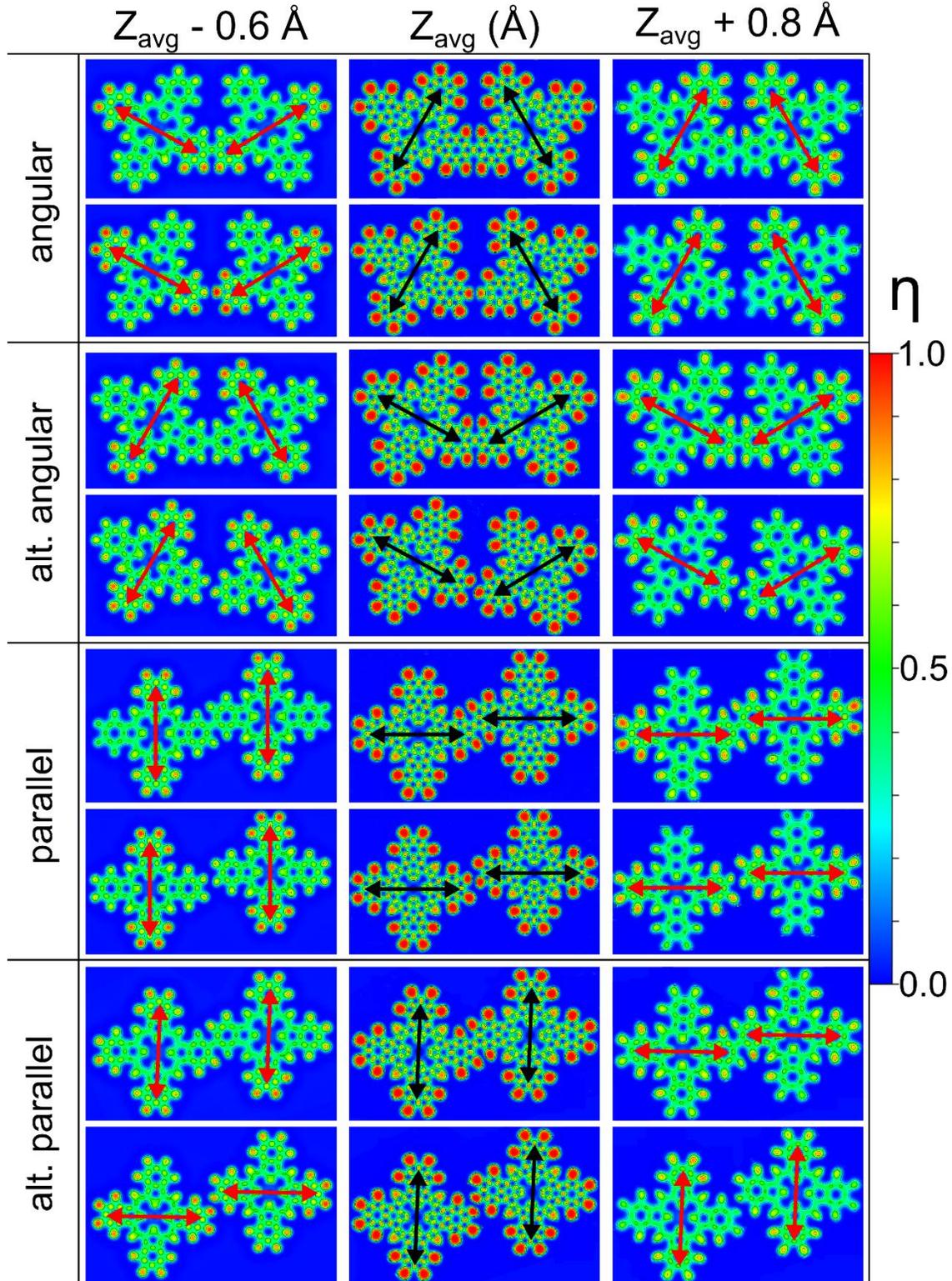

FIG. S4. 2-fold symmetry with ELF slices on bonding and non-bonding CuPc-CuPc geometries. Three slices (0.6 Å below the average z-height, average z-height, and 0.8 Å above the average z-position) demonstrate how different lobes are affected by the substrate orientation. Black arrows show closed packed directions on Cu(111) and red arrows point across lobes with higher η values.

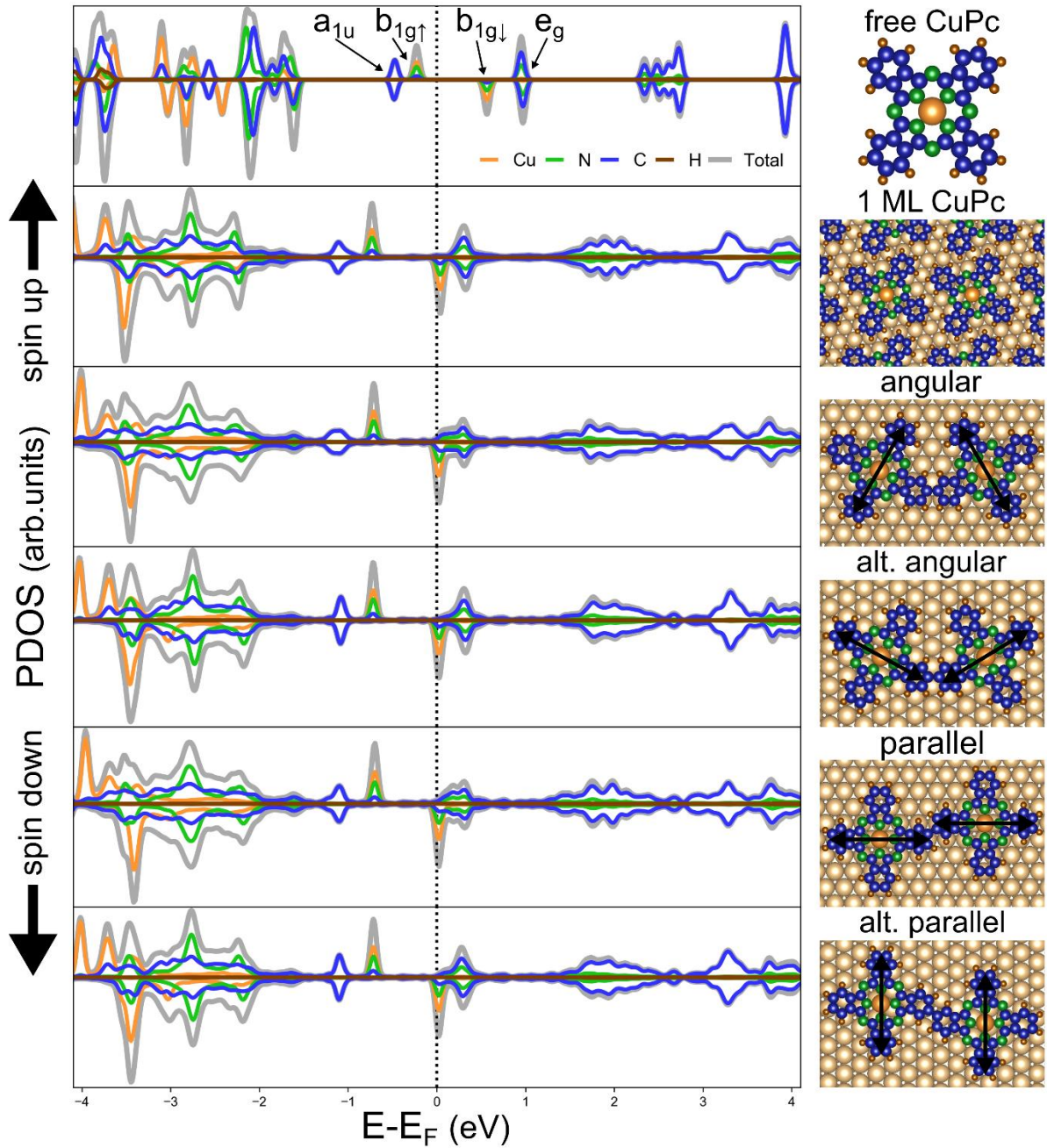

FIG. S5. PDOS plots of CuPc and bonded CuPc in a variety of configurations. Free CuPc shows well-defined orbitals. All configurations of CuPc on Cu(111), including bonded CuPc, show very similar plots. Black arrows on the atomic model show close packed directions of the Cu(111) surface.

## S5. CuPc scattering around Cu(111) surface defects at low concentrations

In Fig. 3(b) of the main text the diffusive background noise does not display any apparent scattering peaks around the surface defects or CuPc clusters. This is primarily because the concentration of non-bonding CuPc is low (0.15 ML mobile CuPc) and that the scattering peak height is small in comparison to the height of immobile CuPc clusters. Fig. S3 displays the same surface twice, were the first order peaks are visible after adjusting the image contrast. The peak width around surface defects and CuPc clusters is ~15 Å, which agrees with previous experimental and Monte Carlo modeling of CuPc scattering around surface defects [19,20]. Given that the number of peaks around surface defects increases with CuPc coverage, only observing one peak at 0.15 ML is consistent with previous experiments and with STM images of $F_{16}$CuPc on Si(111)-(1×1)-Tl at a similar concentration [21].

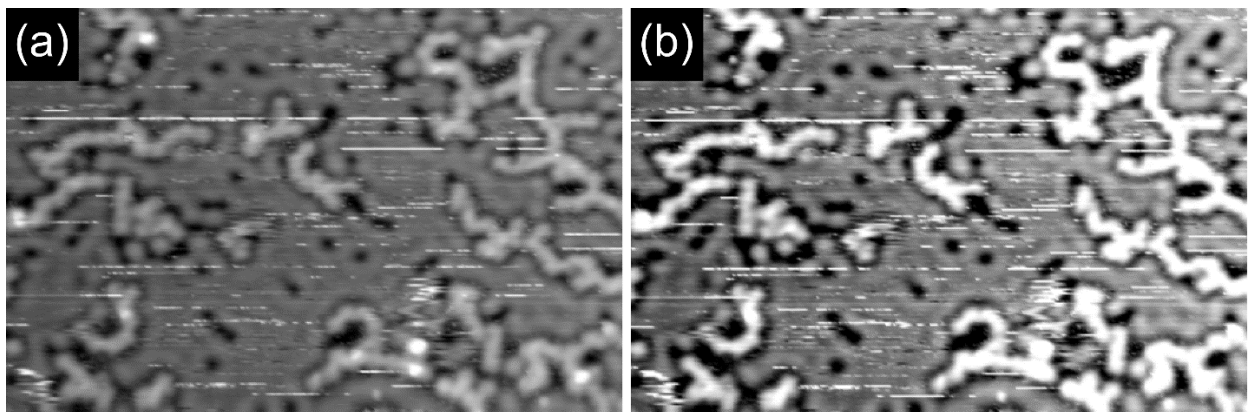

FIG . S6. (a-b) 0.15 ML annealed CuPc with 0.15 ML non-annealed CuPc on Cu(111) before and after contrast enhancement to show the first order scattering peaks around surface defects (+1.7 V, 0.05 nA, 600×400 Å$^2$).

## S6. Estimated diffusion speeds and diffusion coefficients

Typical tip scanning speeds in the STM images varied from ~ 50 to 200 nm/s and typical tip dwell time per pixel varied from 1 to 1.5 ms. From these numbers it is possible to make minimum speed estimations for the observed motion of the CuPc molecules and bonded CuPc clusters. For the case of single CuPc molecules that were imaged as a diffusive background we can estimate a minimum speed on the order of $1.5 \times 10^3$ nm/s based on the assumption that during a 1 ms dwell time at least one diameter of a CuPc molecule (~1.5 nm in size) moves under the tip.

For fast moving bonded CuPc clusters that appear as fuzzy streaks (such as those in denoted by white arrows in Fig. 3 of the main text), we can estimate a minimum speed by measuring the shifts of the feature from consecutive line scans. This procedure measure average speed in fast scanning direction and finds that these clusters move at roughly $1.2 \times 10^0$ nm/s. The speeds were calculated by fitting line scans with Gaussian function (and a linear background if needed) to measure the displacements in the slow scanning direction and dividing by the line scan time, which was ~ 1 s. It should be noted that the streaky clusters chosen for analysis had several (more than 10) line scans in a row with distinguishable feature. Many streaky clusters had discontinuous streak, which is evidence of faster and/or freer motion. Thus, $1.2 \times 10^0$ nm/s is a lower bound for the diffusion speed. This seed is, however, significantly lower than the speed of the single CuPc molecule, indicating that bonding, and surface defects really do affect the way that the clusters moves on the surface. It was also found deposition of additional single CuPc on the surface did not reduce the measured speed of streaky clusters.

Estimation for the speed of slowly moving clusters (typically larger clusters) was performed by measuring the displace of clusters from image to image. These clusters moved with average speed on the order of $10^{-2}$ to $10^{-3}$ nm/s. As this clusters were relatively stable during scanning is was also possible to correlate the speed of the cluster to its size (in number of CuPc molecules). While there was a weak correlation (larger clusters tending to move less), it is very difficult to include important aspects such as the shape of the cluster and the local density surface defects that hindered the motion. Thus, no quantitative statement can be made. It should be noted that this analysis does not account for Brownian motion of the cluster in between the images and therefore is only a roughly estimate of a lower bound for the speed of diffusion.

The diffusion coefficient D can be estimated using the relationship between the mean square displacement $\langle r^2 \rangle$ for Brownian motion;

$$\langle r^2 \rangle = 2nDt \qquad (3)$$

where n is the dimension of the Brownian motion, and t is the time difference between positions. For single CuPc, a displacement of 1.5 nm in 1 ms gives D ~ $10^{-11}$ cm$^2$/s. For streaky clusters moving in between line scans, one-dimensional Brownian motion gives D ~ $10^{-14}$ cm$^2$/s. For displacements between images, 2D Brownian motion gives D ~ $10^{-17}$ cm$^2$/s.